\documentclass[prl,twocolumn,showpacs,superscriptaddress,floatfix]{revtex4}
\usepackage{graphicx}

\def\nbZ{{\mathchoice {\hbox{$\sf\textstyle Z\kern-0.4em Z$}} 
{\hbox{$\sf\textstyle
Z\kern-0.4em Z$}} {\hbox{$\sf\scriptstyle Z\kern-0.3em Z$}} 
{\hbox{$\sf\scriptscriptstyle
Z\kern-0.2em Z$}}}}

\begin{document}

\title{Quantum dynamics in high codimension tilings: from quasiperiodicity to disorder}

\author{Julien Vidal}
\address{Groupe de Physique des Solides, CNRS UMR 7588, 
Universit\'{e}s Paris 6 et  Paris 7,
2, place Jussieu, 75251 Paris Cedex 05 France}

\author{Nicolas Destainville}
\address{ Laboratoire de Physique Th\'eorique, 
CNRS/Universit\'{e} Paul Sabatier,
118 Route de Narbonne, 31062 Toulouse Cedex 04 France}

\author{R\'emy Mosseri}
\address{Groupe de Physique des Solides, CNRS UMR 7588, 
Universit\'{e}s Paris 6 et  Paris 7,
2, place Jussieu, 75251 Paris Cedex 05 France}

\begin{abstract}

We analyze the spreading of wavepackets in two-dimensional quasiperiodic and random
tilings as a function of their codimension, i.e. of their topological complexity.
In the quasiperiodic case, we show that the diffusion exponent that characterizes the
propagation decreases when the codimension increases and goes to 1/2 in the high
codimension limit.  By constrast, the exponent for the random tilings is independent of
their codimension and also equals 1/2. This shows that, in high codimension, the
quasiperiodicity is irrelevant and that the topological disorder leads in every case,
to a diffusive regime, at least in the time scale investigated here.
 
\end{abstract}


\maketitle

It is now well established that quasiperiodic order has a strong 
influence on the quantum dynamics of wavepackets. Indeed, the nature of the
eigenstates in  quasiperiodic systems, which are neither spatially extended (as in
periodic systems) nor  localized (as in disordered systems) but {\em critical},
is often responsible for a sub-ballistic motion.  Although most of the studies
about this anomalous diffusion concern one-dimensional systems such as the
Fibonacci or the Harper chain, there has also been a great interest for
the, more physical, higher-dimensional
ones \cite{Passaro_Octo,Zhong,Yuan,Triozon_Rauzy,Vidal_ICQ7}. However, the  parameters
that determine the characteristics of the long time dynamics, such as the diffusion
exponent $\beta$, remains misunderstood.

\begin{figure}[ht]
\includegraphics[width=88mm]{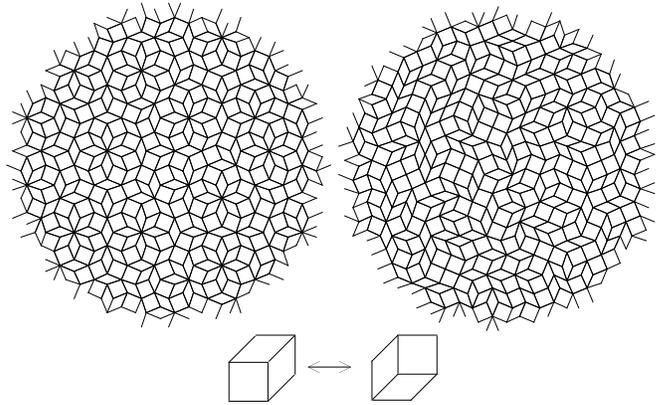}
\caption{ Left: Quasiperiodic tiling built with the grid method with
$D=4$. Right: Random tiling obtained from the quasiperiodic tiling by
elementary flips (center). Such transformations conserve the
st{\oe}chiometry of the tiles but destroy quasiperiodicity and
change the coordination number of the sites.}
\label{tilings}
\end{figure}

In this paper, we investigate the quantum dynamics of wavepackets in two-dimensional
quasiperiodic tilings built with De Bruijn grid method \cite{deBruijn1,deBruijn2}. This construction
allows us to change easily the codimension of the structures that determines the number
of different tiles and thus, in a sense, its complexity. The codimension reveals itself
as a fundamental quantity for the dynamics since the diffusion exponent decreases when
the codimension increases  and converges towards $1/2$ in the high codimension
limit.  We also analyze the influence of phason flips that consists in turning the quasiperiodic tiling
into a random one without changing the st{\oe}chiometry of the tiles (see Fig. \ref{tilings}). We show
that contrary to what was claimed in Ref.
\cite{Passaro_Octo} for the octagonal tiling, the phason disorder do not fasten the
spreading but, always slow it down towards a diffusive regime. 

The De Bruijn grid algorithm \cite{deBruijn1,deBruijn2} relies on two steps.  First, we define $D$
families of $N_l$ equally spaced parallel lines, rotated one from the other by an angle $2\pi/D$.
These  $D \times N_l$ lines form  a grid and separate different connected regions of the plane called
cells of the decomposition. Second, we dualize this grid by associating a tiling vertex to each cell
and connecting any vertices associated to two cells separated by a line. 
%
%
%
%

This algorithm is simpler to implement in two dimensions than the standard Cut-and-Project
\cite{KD,KKL,Elser_random_tiling} method where tilings are generated from high-dimensional
spaces. In this latter approach, the codimension of the tiling is defined by the
difference between the dimension $D$ of the initial lattice in which the points are selected,
and the dimension $d$ of the tiling. 
For two-dimensional tilings, the Cut-and-Project method requires the analyzis of a
$(D-2)$-dimensional (perpendicular) space whose complexity increases very quickly with $D$ so
that, in practice, this method is not tractable beyond $D=5$. 
By contrast, in the grid method, a tiling of codimension $(D-2)$ is straightforwardly
obtained by choosing $D$ families of lines. Note that for $D=2$, the grid method builds the
usual square lattice, and, for $D=3$, it builds the well-known (periodic) dice lattice that
displays interesting properties when embedded in a magnetic field
\cite{Vidal_Cages_big}. Nonperiodic (quasiperiodic) tilings are thus
obtained for $D\geq 4$. The number of lines $N_{l}$ determines the size of the tiling
since the number of sites is $N_{s} \simeq N_{l}^2 D (D-1)/2$. 
However, with this method, different tilings can be obtained depending on the relative
position of each line families. Here, we have focussed on a generic family of tilings and
checked that our results were weakly sensitive to this dephasing choice. In addition, to
avoid spurious effects due to fixed boundary conditions \cite{Destainville} ($N_{l}$ is
finite), we have isolated a circular central cluster of about $10^6$ sites, which is
quasiperiodic.

The dynamics is given by a standard tight-binding Hamiltonian:
%
%
\begin{equation} H = -\sum_{\langle  i,j\rangle} t_{ij}  \, |i 
\rangle \langle  j|
\label{Hamiltonian}
\mbox{,}
\end{equation}
%
%
where $|i\rangle$ denotes a localized orbital on site $i$ and where 
the hopping term $t_{ij}$ equals to 1 if $i$ and $j$ are nearest neighbors and $0$ 
otherwise.  Of course, the time evolution of any wavepacket is directly obtained by 
diagonalizing $H$, but unfortunately, exact diagonalizations are restricted to rather
small system size  ($\sim 10000$ sites). Since we are interested in the long time behavior
of the dynamics that requires large systems, we have used an approximate method, the Second
Order Differencing Scheme  \cite{Numerique}, which consists, for a given initial state
$|\psi(0)\rangle$, in writing:
%
%
\begin{equation} |\psi(t+\Delta t)\rangle =|\psi(t-\Delta t)\rangle 
-2i \Delta t H
|\psi(t)\rangle
\mbox{.}
\end{equation}
%
%
Practically, a time step $\Delta t=0.05$ is sufficient to get a good
accuracy and all our results have been obtained with this value. Note
that despite the very low order of this development for the evolution
operator $e^{iHt}$, this algorithm is actually an very efficient tool
(see for example \cite{Vidal_ICQ8}).

The time evolution of a wavepacket can be characterized by several 
observables. Here, we focus
on the mean square spreading defined by:
%
%
\begin{equation}
\Delta R^2(t)=
\langle \psi(t)|\hat R^2|\psi(t)\rangle - [\langle \psi(t)|\hat 
R|\psi(t)\rangle]^2
\mbox{,}
\end{equation}
%
%
where $\hat R$ is the position operator, the origin
being taken at the center of the cluster. The diffusion exponent $\beta$ is defined by the
long time behavior of $\Delta R^2(t) \sim t^{2\beta}$. {\it A priori},
nothing prevents from getting a different asymptotic regime but, actually, $\Delta R^2$
often seems to be rather well described by a power-law at large time.  The case $\beta=1$
obtained for periodic potential defines a ballistic propagation whereas $0<\beta<1$, observed
in many quasiperiodic systems, corresponds to a sub-ballistic spreading.  Note
that the diffusive motion ($\beta=1/2$) is obtained in several situations among which the
three-dimensional Anderson model in the metallic part of the spectrum. A crucial issue is to
understand the parameters which determine the value of this exponent.

As already noticed in the octagonal tiling \cite{Passaro_Octo} for
wavepackets initially localized on a single site, $\beta$ strongly
depends on this site. In the Labyrinth tiling \cite{Yuan} and in the
generalized Rauzy tilings \cite{Triozon_Rauzy} where energy-filtered
wavepackets have been studied, $\beta$ has also been shown to be
energy-dependent. In the present work, we do not claim to give a
precise characterization of the diffusion exponent for each tiling but
we wish to investigate the behavior of $\beta$ with respect to the codimension.
Therefore, we have chosen to consider random phase states initially spread over a
finite portion of the cluster (typically $1\%$). This leads to an exponent which is
``averaged" over the total density of states. We have considered several configurations of the
phases and several initial radius, and checked that although $\beta$
is different from one state to another, it fluctuates of less than
$5\%$. 
In addition, since we have finite size systems with open boundary conditions, 
the propagation must be stopped when the wavepacket reach the boundaries.
As a criterion for this maximum time $t_{\tiny \mbox{max.}}$, we have chosen the time
for which the presence probability integrated over all the border
sites reaches $1/ N_{s}$. Typically, this leads to $t_{\tiny \mbox{max.}} \simeq
300$. As usual, we can never ensure that the real asymptotic regime  (if it exists~!) is reached
but this time scale certainly gives some hints concerning the long time behavior.

\begin{figure}[ht]
\includegraphics[width=80mm]{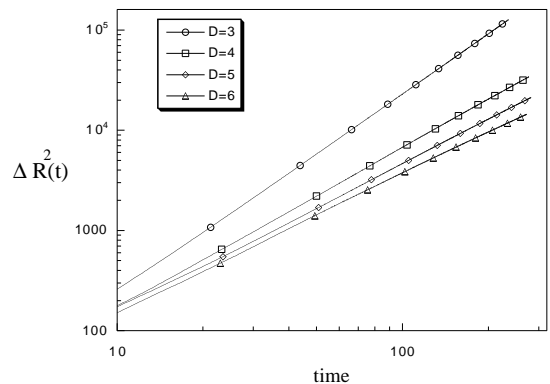}
\caption{Log-log plot of $\Delta R^2(t)$ for low codimension 
quasiperiodic tilings.}
\label{lowcodim}
\end{figure}

The mean square spreading $\Delta R^2(t)$ is displayed in Fig. \ref{lowcodim} 
for quasiperiodic tilings with
$D\in[3,6]$. The dice lattice $D=3$ is given as a reference for a 
ballistic motion ($\beta=1$). For low codimension tilings, we observe a superdiffusive regime
with  an exponent that decreases when the codimension increases.
%
%
%
%

More precisely, we get  $\beta=0.77(1)$ for the octagonal tiling $(D=4)$, $\beta=0.71(1)$ for
the Penrose tiling  $(D=5)$, and $\beta=0.65(1)$ for the dodecagonal tiling $(D=6)$.
Nevertheless, we emphasize that even  for these codimensions, the value of the diffusion
exponent is not constant over a wide time  range and actually slightly decreases when the time
increases. This leads us to conclude  that either the asymptotic regime is not reached, or the
long time behavior is not ``strictly"  power-law like.

For higher codimension, the situation is more complicated: no power-law regime
can be clearly distinguished since $\beta$ strongly depends on the time range and decreases
when the time increases.
%
%
\begin{figure}[ht]
\includegraphics[width=80mm]{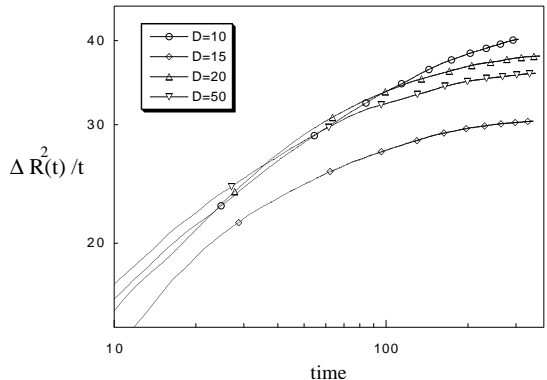}
\caption{Log-log plot of $\Delta R^2(t)/t$ for high codimension 
quasiperiodic tilings.}
\label{highcodim}
\end{figure}
%
%

To determine whether it converges towards a asymtotic value (at large  time) or
not, we have focussed on the behavior of $\Delta R^2(t)/t$ which gives us an
indication on the  super- or sub-diffusive character of the propagation.  This is
an important issue since in such two-dimensional quasiperiodic systems, no sub-diffusive
propagation has ever been reported. As it can be seen in  Fig. \ref{highcodim},
although no power-law behavior can be extracted in the time range investigated, a
strong  indication of a diffusive asymptotic regime 
(${\displaystyle \lim_{t\rightarrow \infty}} \Delta R^2(t)/t=  \mbox{Cte}$) 
is provided.

These results show that the codimension is an important parameter for
the wavepacket propagation.  Indeed, noting that for the two-dimensional
generalized Rauzy tiling \cite{Vidal_ICQ8} $(D=3)$ one obtains an exponent
$\beta=0.95(1)$, it is clear that the higher the codimension, the lower $\beta$. 
This can be easily understood in terms of the complexity of the structure for the following
reasons. An important characteristic of quasiperiodic tilings is the so-called
repetitivity property initially studied in two dimensions by Conway. 
In the class of tilings considered here, it states that for any local environment of linear
typical size $L$, a similar environment can be found at a distance $\xi\sim L^z$ with 
$z \geq 1$. In the simplest case $D=4,5$ one has $z=1$. 
Moreover, when the codimension increases, the number of possible edge orientations 
at each node is lower or equal to $2D$ and for a given $L$, $\xi(L)$ also grows. 
Consequently, the specific effects of the quasiperiodic order are expected to decrease when
the codimension increases.  Note that in periodic system, $\xi$ is a constant and in disordered systems,
it grows exponentially with $L$. 

Nevertheless, these tilings are quasiperiodic and this special order is responsible, at
least in low codimension, for a superdiffusive regime ($\beta >1/2$). 
For higher codimension the quasiperiodicity becomes irrelevant and the exponent
$\beta$ seems to converge towards $1/2$ which is the value expected
for a weakly disordered system in two dimensions (for example a periodic lattice
with random impurities or random magnetic fluxes \cite{Kawarabayashi},
or a disordered quasiperiodic Fibonacci lattice
\cite{Roche_Fibo_desordre}) where such a regime is expected when
$\sqrt{\Delta R^2}$ is lower than the localization
length. Unfortunately, since the asymptotic regime is not reached with
the system size studied here, we cannot estimate the codimension
dependence of the diffusion constant. However, the results shown in
Fig. \ref{highcodim} indicate that this constant does not seem to simply
decrease when the codimension increases.

If this analysis is correct, destroying quasiperiodicity in these
tilings should also lead to a diffusive regime. To check this
assumption and to corroborate our previous statement, we have
studied the quantum dynamics in random tilings obtained from the
quasiperiodic tilings by making elementary phason flips (about
$10^{11}$) as shown in Fig. \ref{tilings} for $D=4$. 
The random flip Markov chain has been studied into detail in the cases $D=3$ and
$D=4$~(see \cite{Destain02}),  and the number of flips needed to get a
random tiling uncorrelated to the initial quasiperiodic tiling is indeed of order $N_s^2$.
Furthermore, this number has been computed numerically in higher codimension tilings and
is, in fact, always smaller than $N_s^2/2$ \cite{WDMB}. In the present case, this leads to an upper
bound of $5.10^{11}$ flips. However, we have checked for $D=4$, that the spreading in tilings
with $10^{11}$ and $5.10^{11}$ flips have essentially the same diffusion exponent.
%
%
\begin{figure}[ht]
\includegraphics[width=80mm]{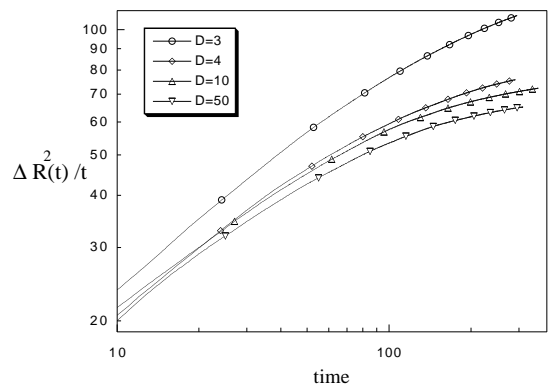}
\caption{Log-log plot of $\Delta R^2(t)/t$ for random tilings.}
\label{random}
\end{figure}
%
%

\begin{figure}[ht]
\includegraphics[width=80mm]{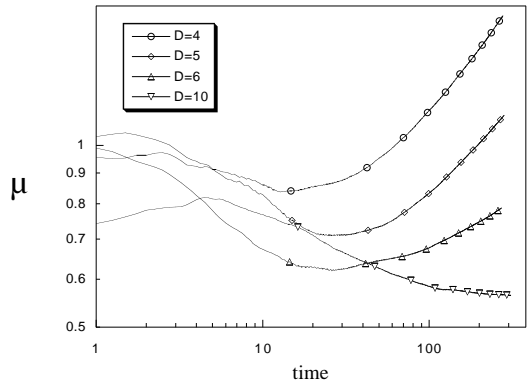}
\caption{Log-log plot of 
$\mu=\Delta R^2_{\tiny \mbox{quasi.}}/ \Delta R^2_{\tiny \mbox{random}}$ as a
function of time.}
\label{quasirandom}
\end{figure}

As shown in Fig. \ref{random}, although no power-law regime can be 
clearly extracted, the propagation is very similar to those displayed in Fig. \ref{highcodim}
for high codimension quasiperiodic tilings. For the octagonal tiling ($D=4$), this result
refutes the conclusion of  Passaro {\it et al.} claiming that the diffusion exponent increases
when a phason disorder is  introduced. This contradiction is due to the short time scale
investigated in Ref. \cite{Passaro_Octo}. Indeed, we  also observe that for $t\lesssim 20$,
the propagation is faster in the $D=4$ random tiling  than in the $D=4$ quasiperiodic tiling.
However, at larger times, the motion in the random tiling  seems to become diffusive whereas,
in the octagonal tiling, it remains superdiffusive with $\beta=0.77(1)$ \cite{ZZZ}. 
We have also observed a similar crossover for higher codimension $(D=5,6)$ (see
Fig. \ref{quasirandom}). 
%
%
%
%

A possible explanation of these two regimes is that, at short times,
the local quasiperiodic order is likely res\-ponsible for coherent quantum
interferences (absent in a random tiling) that slow down the dynamics. In this
regime, the topological disorder thus favors, in a sense, the propagation. At
large times, the influence of the quasiperiodicity remains the same
but a weak localization process becomes dominant for the random tiling
and leads to a diffusive regime. Of course, for larger codimension, the situation
is different since the asymptotic regime is likely the same both in the
quasiperiodic and in the random case. Thus, one expects 
$\mu=\Delta R^2_{\tiny \mbox{quasi.}}/ \Delta R^2_{\tiny \mbox{random}}$ to
become constant at large time and it is actually the case (see Fig. \ref{quasirandom} for $D=10$).

This study raise interesting questions concerning the nature of the spectrum in
quasiperiodic and random tilings. A simple way to characterize the spectral measure of a
state $|\psi\rangle$ is to compute the averaged autocorrelation function
%
%
\begin{equation}
C(t)={1\over t} \int_0^{t} dt' P(t') \sim t^{-\alpha}
\mbox{,}
\end{equation}
%
%
where $P(t)=|\langle \psi(0)|\psi(t)\rangle|^2$. If $\alpha=1$, the spectral measure is
absolutely  continuous, if $\alpha=0$, it  is pure point, and if $0<\alpha<1$ it is  singular
continuous. As previously, for a random phase state, $\alpha$ gives an average information
over the total density of states. Unfortunately, in the tilings studied here, there exist
strictly localized states known as confined states due to specific local environment
(see Ref. \cite{Kohmoto_Sutherland,Arai2} for the case $D=5$).  These states eventually leads
to $\alpha=0$ provided the initial random phase state has a nonzero overlap with them. 
The analyzis of the spectral measure with respect to the codimension thus requires to get
rid of these confined states. Further investigations are needed to clarify this point which is 
certainly of great interest in the understanding of the electronic properties of quasiperiodic systems.

\acknowledgments
We would like to thank M. Baake and M. Duneau for fruitful remarks about the repetitivity property
of quasiperiodic tilings.


\begin{thebibliography}{10}

\bibitem{Passaro_Octo}
B. Passaro, C. Sire, and V.~G. Benza, Phys. Rev. B {\bf 46},  13751  (1992).

\bibitem{Zhong}
J.~X. Zhong and R. Mosseri, J. Phys. C {\bf 7},  8383  (1995).

\bibitem{Yuan}
H.~Q. Yuan, U. Grimm, P. Repetowicz, and M. Schreiber, Phys. Rev. B {\bf 62},
  15569  (2000).

\bibitem{Triozon_Rauzy}
F. Triozon, J. Vidal, R. Mosseri, and D. Mayou, Phys. Rev. B {\bf 65},  220202
  (2002).

\bibitem{Vidal_ICQ7}
J. Vidal and R. Mosseri,   Proceedings of ICQ7, edited by F. G\"ahler, P. Kramer, H.~R. Trebin,
  and K. Urban (Elsevier, Switzerland, 2000), Vol.~A294-A296, p.\ 572.

\bibitem{deBruijn1}
N.~G. de~Bruijn, Proc. Konink. Ned. Akad. Wetensch. {\bf A 43},  84  (1981).

\bibitem{deBruijn2}
N.~G. de~Bruijn, J. Phys. France {\bf 47},  C3  (1986).

\bibitem{KD}
M. Duneau and A. Katz, Phys. Rev. Lett. {\bf 54},  2688  (1985).

\bibitem{KKL}
P.~A. Kalugin, A.~Y. Kitaev, and L.~S. Levitov, J. de Phys. (Paris) Lett. {\bf
  46},  L601  (1985).

\bibitem{Elser_random_tiling}
V. Elser, Phys. Rev. Lett. {\bf 54},  1730  (1985).

\bibitem{Vidal_Cages_big}
J. Vidal, P. Butaud, B. Dou\c{c}ot, and R. Mosseri, Phys. Rev. B {\bf 64},
  155306  (2001).

\bibitem{Destainville}
N. Destainville, J. Phys. A {\bf 31},  6123  (1998).

\bibitem{Numerique}
{C. Leforestier {\it et~al.}}, J. Comp. Phys. {\bf 94},  59  (1991).

\bibitem{Vidal_ICQ8}
J. Vidal and R. Mosseri,  Proceedings of ICQ8, cond-mat/0209037, unpublished.

\bibitem{Kawarabayashi}
T. Kawarabayashi and T. Ohtsuki, Phys. Rev. B {\bf 51},  10897  (1995).

\bibitem{Roche_Fibo_desordre}
S. Roche and D. Mayou, Phys. Rev. Lett. {\bf 79},  2518  (1997).

\bibitem{Destain02}
N. Destainville, Phys. Rev. Lett. {\bf 88},  030601  (2002), and references therein.

\bibitem{WDMB}
M. Widom, N. Destainville, R. Mosseri, and F. Bailly, unpublished.

\bibitem{ZZZ}
For $D=3,4$, it is possible to have an explicit form for the sites coordinates
  thanks to a special numbering. Thus, for the Rauzy tiling and the octagonal
  tiling, we have checked that $\beta$ remains the same up to $1\%$ for tilings
  with about $8.10^6$ sites!

\bibitem{Kohmoto_Sutherland}
M. Kohmoto and B. Sutherland, Phys. Rev. B {\bf 34},  5043  (1986).

\bibitem{Arai2}
M. Arai, T. Tokihiro, T. Fujiwara, and M. Kohmoto, Phys. Rev. B {\bf 38},  1621
   (1987).

\end{thebibliography}
\end{document}